# Grain boundary segregation of interstitial and substitutional impurity atoms in alpha-iron

M. Rajagopalan[1], M.A. Tschopp[2], K.N. Solanki[1*]

[1]School for Engineering of Matter, Transport, and Energy, Arizona State University, Tempe, AZ 85287, USA
[2] U.S. Army Research Laboratory, Weapons and Materials Research Directorate, Aberdeen Proving Ground, MD 21005
*Corresponding Author: (480) 965-1869; (480) 727-9321 (fax), E-mail: kiran.solanki@asu.edu

**Abstract**

The macroscopic behavior of polycrystalline materials is influenced by the local variation of properties caused by the presence of impurities and defects. The effect of these impurities at the atomic scale can either embrittle or strengthen grain boundaries within. Thus, it is imperative to understand the energetics associated with segregation to design materials with desirable properties. Here, molecular statics simulations were employed to analyze the energetics associated with the segregation of various elements (helium, hydrogen, carbon, phosphorous, and vanadium) to four <100> ($\Sigma 5$ and $\Sigma 13$ GBs) and six <110> ($\Sigma 3$, $\Sigma 9$, and $\Sigma 11$ GBs) symmetric tilt grain boundaries in alpha-Fe. This knowledge is important for designing stable interfaces in harsh environments. Simulation results show that the local atomic arrangements within the GB region and the resulting structural units have a significant influence on the magnitude of binding energies of the impurity (interstitial and substitutional) atoms. This data also suggests that the site-to-site variation of energies within a boundary is substantial. Comparing the binding energies of all ten boundaries shows that the $\Sigma 3(112)$ boundary possesses a much smaller binding energy for all interstitial and substitutional impurity atoms among the boundaries examined here. Additionally, based on the Rice-Wang model, our total energy calculations show that V has a significant beneficial effect on the Fe grain boundary cohesion, while P has a detrimental effect on grain boundary cohesion, much weaker than H and He. This is significant for applications where extreme environmental damage generates lattice defects and grain boundaries act as sinks for both interstitial and substitutional impurity atoms. This methodology provides us with a tool to effectively identify the local as well as the global segregation behavior which can influence the GB cohesion.

## 1. Introduction

Increasing global demand for safer, energy-efficient, bio-compatible systems for biomedical, transportation, and safety applications requires developing new materials with tuned interface structures [1]. Because the mechanical behavior and fracture of polycrystalline materials is often driven by grain boundaries and their underlying structure [2–4], a fundamental understanding of the relationship between the grain boundary structure and associated properties is important to develop interface-dominant materials. Research has shown that both the macroscopic degrees of freedom and microscopic local structure affect the physical properties of grain boundaries [5–

12]. The term *grain boundary character* is often used to describe the five degrees of freedom necessary to define a grain boundary. Three degrees of freedom are used to define the misorientation between the two grains and two degrees of freedom are associated with the GB plane. In terms of the microscopic local structure, the translations between adjoining grains are also important, as is the localized dislocation structure of the boundary. Historically, many efforts have focused on developing a method to characterize grain boundaries [13–18] and their influence on the physical properties of polycrystalline materials. These models utilize dislocation arrays, disclinations, and coincident site lattice (CSL) to describe the local structure of grain boundaries. These efforts, in turn, have led to identifying the primary structural elements for symmetric tilt, asymmetric tilt, twist, and twin boundaries at the atomic scale [5,8,19–27]. The term structural unit (SU) has been used to describe the local atomic arrangement at the grain boundary and is associated with both the grain boundary character and its properties (e.g.[9,28–32]). Experimentally, Saylor *et al.* [33] studied the grain boundary character distribution (GBCD) as a function of grain boundary geometry for a commercially pure Al sample. They indicated that boundaries with lower energies and index planes have a higher distribution in the polycrystalline sample. These results also apply to other metals such as nickel (Ni) and copper (Cu). Experimentally, grain boundary structure has been observed using field ion microscopy and high resolution transmission electron microscopy [34–39]. The grain boundary energies can be computed through theoretical formulations and computational methods. Thus, understanding the atomic structure at the grain boundary can provide insight into the grain boundary strength as well as various grain boundary dependent phenomena, such as diffusion and segregation [40–42].

Quantifying how point defects interact with defect sinks, such as grain boundaries, is also important for understanding the strength of interfaces in high radiation and corrosive environments. For instance, during irradiation-induced segregation, the flux of solute and impurity elements is highly coupled with the flux of vacancies and interstitials. As vacancies and interstitials tend to diffuse and bind to microstructural sinks, solute and impurity atoms are spatially redistributed in the vicinity of these sinks [43]. The net result is an accumulation or a depletion of elements at these defect sinks, which can have deleterious effects on polycrystal properties [44]. Atomistic and electronic simulations are increasingly being utilized as tools for investigating such fundamental mechanisms associated with segregation and binding behavior. It has been shown that impurity segregation to grain boundaries can have a profound effect on the mechanical behavior in polycrystals, i.e., a significant beneficial effect [45–47] or a significant detrimental effect [41,48–52]. For example, Solanki *et al.* found that certain H defects are favored at $\alpha$-Fe grain boundaries and that these species affect the grain boundary cohesive strength [10]. On the other hand, Yamaguchi [47] has shown that the segregation of boron and carbon to grain boundaries can actually be beneficial by strengthening grain boundary cohesion. These results indicate that the segregation behavior of these elements plays an important role in grain boundary embrittlement or strengthening behavior. Moreover, the grain boundary character can influence the segregation behavior. Recently, Tschopp *et al.* [11,53] used molecular statics simulations with various Sigma ($\Sigma$) grain boundaries to show how the local grain boundary structure and the macroscopic grain boundary character affects the sink strength for vacancy and self-interstitial point defects. In addition to quantifying the binding energetics and the absorption length scale of point defects, this work found that there is an energetic preference for self-interstitial atoms to preferentially bind to grain boundaries over vacancies, in agreement with other recent studies [54]. This work has been further extended to systematically quantify the interactions of point defects, carbon, hydrogen and helium with Fe grain boundaries [10,11,55,56]. These studies provide a generalized framework for exploring how grain boundary character can affect the segregation of various elements to a grain boundary. Here, we used this framework to explore the relationship between the local grain boundary structure and the segregation energetics using both interstitial and substitutional impurity atoms in alpha-iron. Of particular interest was showing whether different impurity atoms have beneficial or weakening effects on the cohesion in ten low-$\Sigma$ grain boundaries, which represent a range of boundaries observed in polycrystalline iron experimentally. The paper is laid out as follows. Section 2 describes the methodology and GB generation procedures used in our study. Section 3 describes several interesting observations: (1) the grain boundary local arrangements and resulting structural units have a significant influence on the magnitude of binding energies of impurity atoms, and the site-to-site variation within a boundary is substantial, (2) a comparison of the binding energies of all ten boundaries shows that the $\Sigma 3(112)$ boundary possesses a much smaller binding energy for all impurity atoms (interstitial and substitutional), (3) for all grain boundaries examined here, there were atoms lying symmetrically along the grain boundary plane that had binding energies for all impurity atoms close (or even lower, in some cases) to bulk values, i.e., these grain boundary sites are unfavorable to act as a sink for impurity atoms, (4) in most grain boundaries examined here, the vacancy binding energies approach bulk values within 10 Å from the grain boundary center plane, and (5) based on the Rice-Wang model [57], our total energy calculations show that V is beneficial for the grain boundary cohesion in Fe, while P is detrimental for the grain boundary cohesion.

## 2. Methodology

The effect of grain boundary structure on the segregation behavior of P, V, C, H and He was examined using molecular statics simulations to ten different low $\Sigma$ symmetric tilt grain boundaries (STGBs) with <100> and <110> tilt axes in $\alpha$-Fe. A parallel molecular dynamics code (Large-scale Atomic/Molecular Massively Parallel Simulator, LAMMPS [58]) with semi-empirical embedded atom method (EAM) potentials was used to describe the Fe-H, Fe-V, Fe-P, and Fe-C systems [59–62]. In the case of Fe-He, a modified version of the MOLDY code was used [63,64]. These EAM potentials were parameterized using an extensive database of energies and configurations from DFT calculations and have been used to accurately define different material behaviors, such as surface energies, generalized staking fault energies, etc. (e.g., see Refs. [10,11,55,56,59–62,64]). The cohesive energy of Fe predicted by all the potentials is ~4.013 eV/atom.

The simulation cell used for quantifying the interaction of impurity atoms with the GBs are as follows. The equilibrium 0 K grain boundary structure and energy was calculated using a bicrystal computational cell with three-dimensional (3D) periodic boundary conditions consisting of two grains. The minimum distance between the two periodic boundaries in each computational cell was 12 nm. As with past work [10,11,55,56], an atom deletion criterion, multiple initial configurations, and various in-plane rigid body translations were utilized to accurately obtain an optimal minimum energy GB structure via the nonlinear conjugate gradient energy minimization process.

The generalized framework for exploring how grain boundary character can affect grain boundary segregation of various elements is similar to that described by Tschopp et al. [11] and is shown in Figure 1. The methodology used for adding various impurities, both interstitial and substitutional, to grain boundary sites and quantifying binding energy is as follows. In the initialization step, four <100> and six <110> STGBs are generated using the Mendelev et al. [65] Fe potential. In the test step, a grain boundary is selected and a specific grain boundary site is chosen to substitute either a single interstitial or substitutional atom. Then, molecular statics is used to calculate the segregation and binding energies for that particular site. This process is repeated for all sites within 15 Å of the grain boundary center and for all of the grain boundaries. Last, in the analysis step, the calculated properties are examined to determine the influence of important factors, such as local atomic structure, distance from the grain boundary, and grain boundary character (misorientation angle, Sigma value, etc.) on the segregation and binding energies.

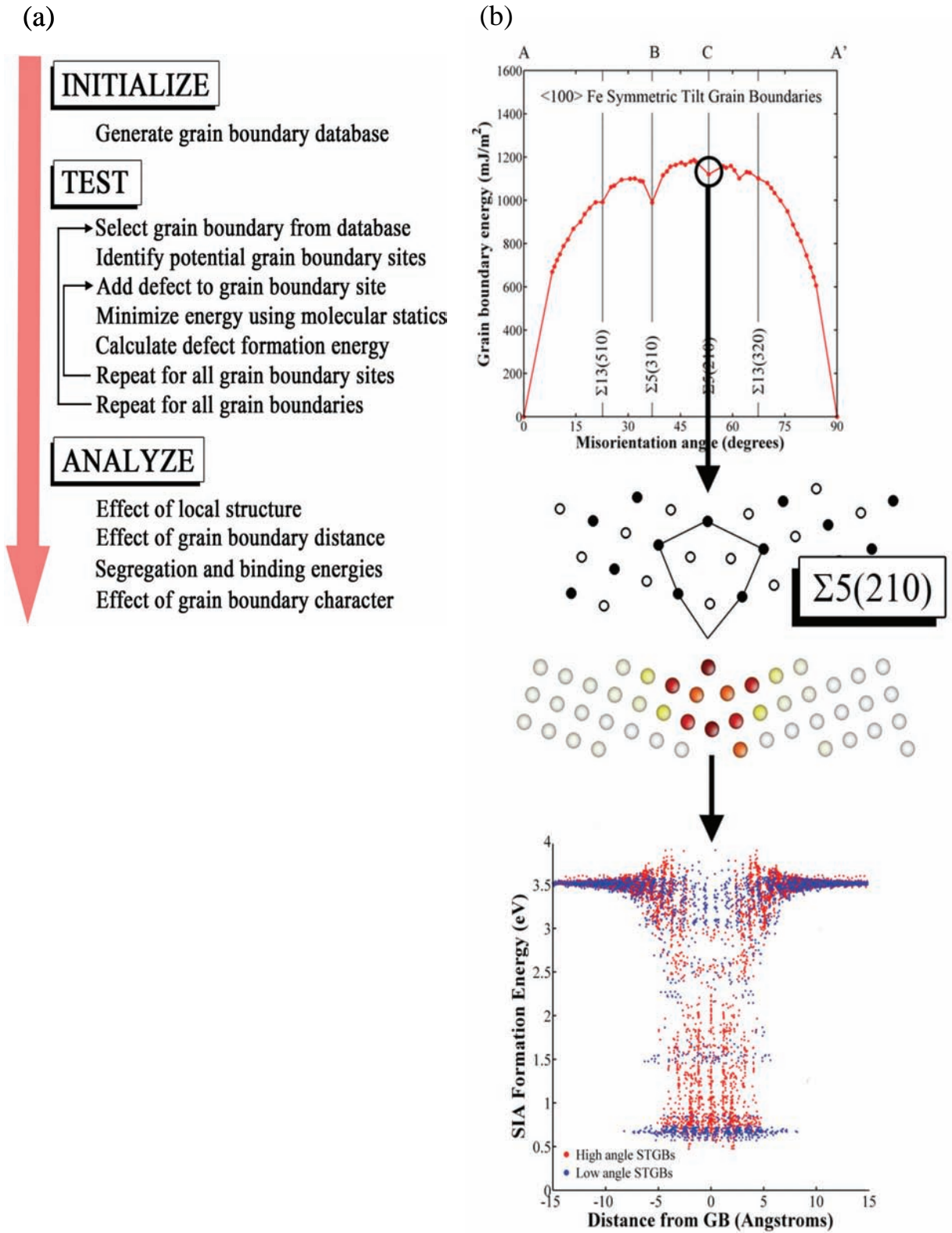

Figure 1. (a) Schematic of the process used to initialize, test, and analyze the segregation of elements to grain boundaries [11], and (b) example of a grain boundary system from which a single grain boundary was selected and then the point defect formation energy of every potential grain boundary site was subsequently tested to build a database of formation energies [11]. The same process was applied herein to quantify element segregation of H, He, P, V, and C to Fe GBs.

The segregation energy (or binding energy) was calculated for impurity atoms as a function of its position at each site within 15 Å of the GB. For each GB structure, an impurity atom was placed at a site α, and the simulation cell was relaxed using the Polak-Ribière [66] conjugate gradient energy minimization process. The total energy of the simulation cell was calculated and the process was repeated for each atomic site within various STGBs. The approach used for calculating the segregation energy is defined elsewhere [10,55,56].

The segregation energy for an impurity atom at the interstitial/substitutional site α was calculated as follows

$$E_{seg}^{\alpha} = (E_{GB}^{\alpha} - E_{GB}) - (E_{bulk}^{\alpha} - E_{bulk}) \tag{1}$$

where $E_{GB}^{\alpha}$ and $E_{GB}$ are the total energies of the GB structure with and without impurity atoms. $E_{bulk}^{\alpha}$ and $E_{bulk}$ are the total energies of a single crystal bulk Fe simulation cell with and without the impurity atom at a particular site. These bulk energies are subtracted in Eq. (1) to remove the bulk's contribution to the energetics of the impurity in the single crystal lattice. For each GB, the position-based segregation energies are obtained and then categorized with respect to positive and negative segregation energies. That is, the preference of the impurity atom to segregate to the GB corresponds to a negative segregation energy value and the preference of the impurity atom to stay in the bulk corresponds to a positive segregation energy value. This convention is a factor of -1 from the binding energy (negative segregation energy is equivalent to a positive binding energy, i.e., segregation is favored).

The GB embrittlement can be assessed from the change in cohesive energy of the GB with and without an impurity. The cohesive energy of a GB is expressed as a difference in the energies between the fractured surfaces and the GB with an impurity atom at a site α. The cohesive energy of a GB in the presence of an impurity atom is given by:

$$2\gamma_{int} = 2\gamma^{\alpha} - \gamma_{gb}^{\alpha} \tag{2}$$

where $2\gamma_{int}$ is the cohesive energy of the two GB surfaces with an impurity atom, $2\gamma^{\alpha}$ is the surface energy of the two free surfaces with an impurity atom, and $\gamma_{gb}^{\alpha}$ is the GB energy with an impurity atom at location α. For a system without an impurity, $2\gamma^{\alpha}$ and $\gamma_{gb}^{\alpha}$ correspond to the surface and GB energy of the pure FS and GB. If the cohesive energy without the impurity is higher than that with the impurity, this implies that decohesion is favored with 'x' number of impurity atoms at energetically favorable sites in the boundary.

The susceptibility to GB embrittlement can also be ascertained by quantifying the difference in segregation energies at the free surface and at the GB. If the segregation energies are lower at the surface, then this indicates an embrittling effect of the impurity atom on the GB [47,52].

## 3. Results and Discussion

## 3.1 Grain Boundary Energy

In the present work, we have focused on how the local grain boundary structure interacts with impurity atoms and how the local atomic environment at the boundary influences the binding energetics of substitutional and interstitial impurity atoms. Here, we investigated the binding behavior of certain elements (H, He, P, V, and C) to a few GB systems in alpha-Fe with tilt axes <100> and <110>. A range of GB structures and energies were generated which served as an input for further calculations. These GBs were both low ($\theta \leq 15°$) and high angle boundaries. Figure 2 shows both the grain boundary energy as a function of misorientation angle for the <110> STGB system and the variation of GB energy as a function of GB geometry in a stereographic triangle representation, which is widely used to represent cubic systems [67]. Here, GB geometry is defined by polar and azimuthal angles, as defined in Ref. [67]. Some of the CSL GBs corresponding to local minima energy for the respective tilt systems are also identified in the plot. For example, the $\Sigma 3(111)\theta=109.47°$ and $\Sigma 9(221)\theta=141.05°$ GBs correspond to shallow cusps in the GB energy plot and the $\Sigma 3(112)\theta=70.53°$ coherent twin boundary [68] shows a more pronounced cusp in GB energy (refer to Figure 2(a)). All the energy values are consistent with energies previously reported [10,11]. The stereographic representation of GB energies for the entire misorientation range helps to identify GBs that show a local minima in GB energies in the energy-misorientation plot (refer to Figure 2(b)). A small subset of such ten GBs corresponding to <100> and <110> tilt systems are then selected for further analysis, which represent a smaller range of boundaries observed in polycrystalline iron experimentally. Table 1 presents a summary of the investigated <100> and <110> STGBs with their corresponding grain boundary energies. Notice that all grain boundaries are low CSL boundaries with the grain boundary energy ranging from 260-1308 mJ/m$^2$. Further details on the GB structures are given in Tschopp *et al.* [11]. Sampling a range of GBs provides a map to identify the local as well as global variation in binding energies with a defect or an impurity which can be used for engineering GBs for favorable properties.

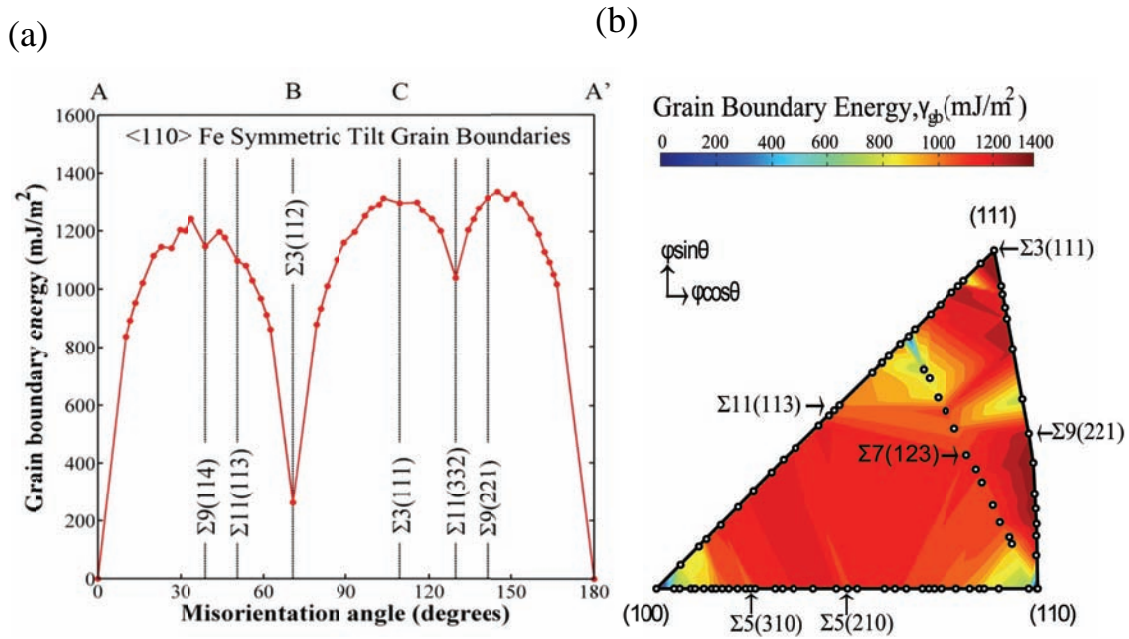

Figure 2. (a) Grain boundary energy for the <110> symmetric tilt grain boundary system. Note the pronounced energy dip for the $\Sigma 3(112)$ coherent twin GB. (b) Contour plot of GB energies

for the three symmetric tilt systems of Fe (<100>, <110>, and <111>) represented using polar and azimuthal angles. The polar and azimuthal angles correspond to the degrees of freedom.

Table 1. Summary of the investigated <100> and <110> STGBs with the corresponding grain boundary energies.

| Sigma (Σ) | Misorientation (Degrees) | GB. Energy (mJ/m$^2$) |
|---|---|---|
| 3{112}<1$\bar{1}$0> | 70.53 | 260 |
| 3{111}<1$\bar{1}$0> | 109.47 | 1308 |
| 5{210}<100> | 36.87 | 1008 |
| 5{310}<100> | 53.13 | 1113 |
| 9{114}<1$\bar{1}$0> | 38.95 | 1169 |
| 9{221}<1$\bar{1}$0> | 141.05 | 1287 |
| 11{113}<1$\bar{1}$0> | 50.48 | 1095 |
| 11{332}<1$\bar{1}$0> | 129.52 | 1207 |
| 13{510}<100> | 22.62 | 1005 |
| 13{320}<100> | 67.38 | 1108 |

3.2 Binding energy

In this section, we investigated the binding energetics as a function of (1) atomic structure (GB type) for P to the four <100> and six <110> STGBs of alpha-Fe, (2) impurity atoms to the two <110> STGBs of Fe namely, the Σ3(112)θ=70.53° and Σ11(332)θ=129.52° GBs, (3) length scale for one interstitial and substitutional impurities, P and H, respectively to the Σ13(320)θ=67.38° STGB.

First, we studied the spatial variation in energetics corresponding to the segregation of an impurity, P to the four <100> and six <110> Fe GBs. This element was chosen because it is known to cause temper embrittlement in steels [69]. The site to site variation of impurity elements can be studied using binding energy calculations which provide an insight into the variation in the binding energy between an impurity atom and a GB along with structural characteristics of the GB. Figure 3 illustrates the spatial variation of binding energies for a substituted P atom in the four selected <100> Fe GBs. This binding energy behavior is very similar to other impurity atoms examined herein. The binding energies are indicative of the binding strength of an impurity to a particular site with positive values favoring a high binding proficiency. There are several observations in this figure: (1) the binding energy value diminishes as the distance from the boundary increases (bulk lattice binding energy is white), (2) majority of the sites show a higher segregation propensity ($E_b^\alpha = -E_{seg}^\alpha$), but there are a few sites within the GB with very low preference to segregate, and (3) the binding energies approach bulk energies within a few atomic layers from the GB.

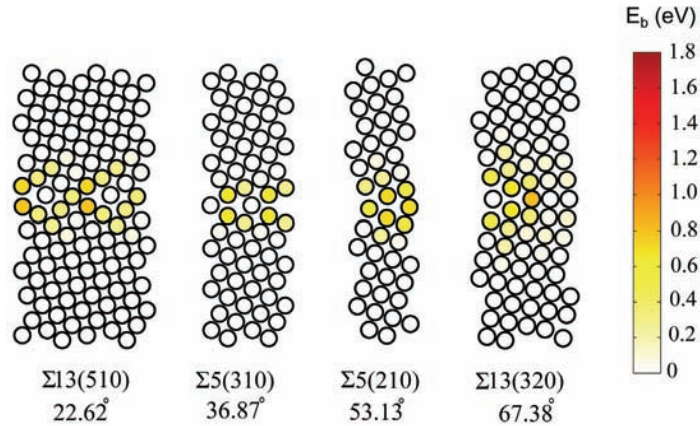

Figure 3. Atomic representation of four <100> GB systems depicting the distribution of the P binding energy. Here, the positive binding energies imply a favorable binding behavior.

This was further extended to assess the binding characteristics of P to the six <110> STGBs. The corresponding plot for binding energy of P to <110> Fe STGBs is depicted in Figure 4. The majority of characteristics indicated for <100> GBs hold true for the <110> GBs. Thus, the present calculations show an energetic preference for a P atom to segregate to the GB region over the bulk. However, there are a few additional characteristics that should be noted. First, the $\Sigma3(112)\theta=70.53°$ GB has a very low binding energy in comparison to the other boundaries. The $\Sigma9$ boundaries have larger binding energies, followed by the $\Sigma11$ boundaries. The $\Sigma3(111)\theta=109.47°$ GB exhibits a maximum binding energy of 1.06 eV at certain sites on either side of the symmetric boundary. In general, the binding energies of <110> GBs are slightly higher than <100> GBs, indicating a preference over the <100> boundaries for a P atom to bind to the substitutional site. These variations can be attributed to the local-arrangement of the atoms at the interface. For each boundary there exists a unique structural arrangement which causes a variation in atomic energies within a GB as well as between different boundaries. Therefore, this structural representation along with site-to-site energy mapping gives an overall detail of how each site can influence the binding/segregation behavior with respect to a particular impurity and how structure-energy can be correlated.

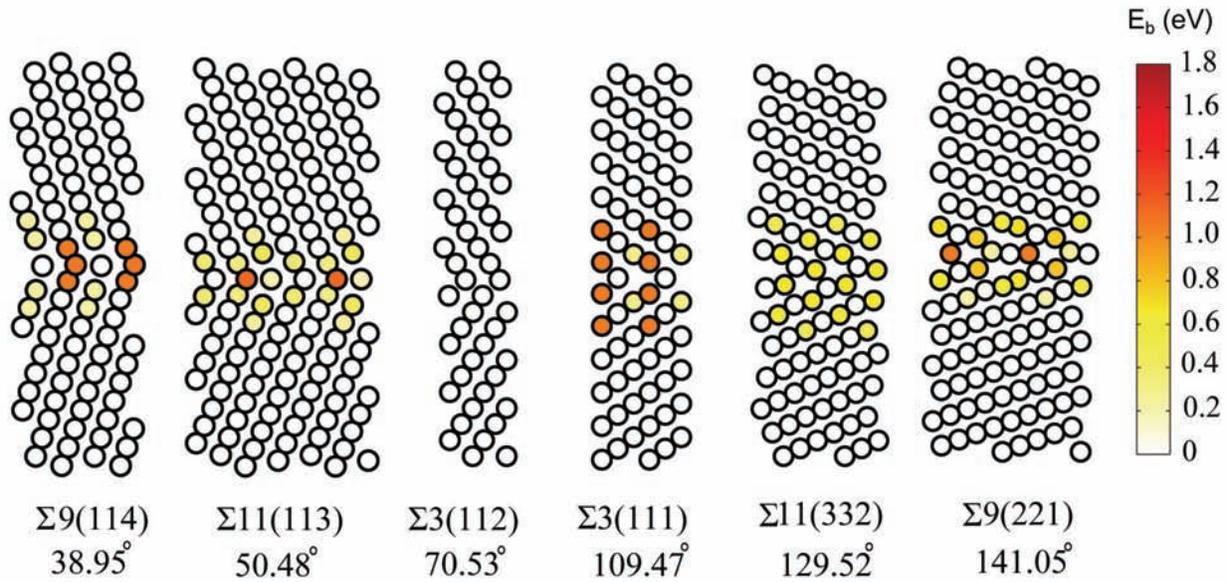

Figure 4. Atomic representation of the six <110> STGBs depicting the distribution of the P binding energy. Positive binding energies imply a favorable binding behavior. Note that the Σ3(112) boundary possesses a much smaller binding energy than other <110> STGBs.

Second, we studied the binding energies at one boundary as a function of different impurities. This helps in qualitative assessment of the propensity of a boundary to the various foreign impurity atoms both at the substitutional and interstitial configurations. Two <110> STGBs are chosen to study the energies at different atomic sites with different substitutional and interstitial impurity atoms. The boundaries are Σ3(112)θ=70.53° and Σ11(332)θ=129.52° GBs. From Figure 4, it is evident that the Σ3(112) GB does not exhibit high binding potency in the case of P, i.e., the grain boundary does not act as an energetically preferable sink for an impurity atom to segregate. This behavior was further expanded to include additional impurity atoms, V substituted atom, and C, H and He interstitial atoms. Figure 5 illustrates the atomic site-energy mapping for the Σ3(112) boundary as a function of impurity atoms. For all the impurity atoms, the binding energy diminishes to 0 eV as we move away from the GB. Additionally, sites within the GB also exhibit binding energies similar to that of bulk. The order of binding behavior at the Σ3(112) GB listed from highest to lowest is Fe-C, Fe-H, Fe-He, Fe-P, and Fe-V respectively. The low binding energies of P and V suggest that the GB sites are not energetically preferable for the substitutional atom to move to when compared to other impurities studied here.

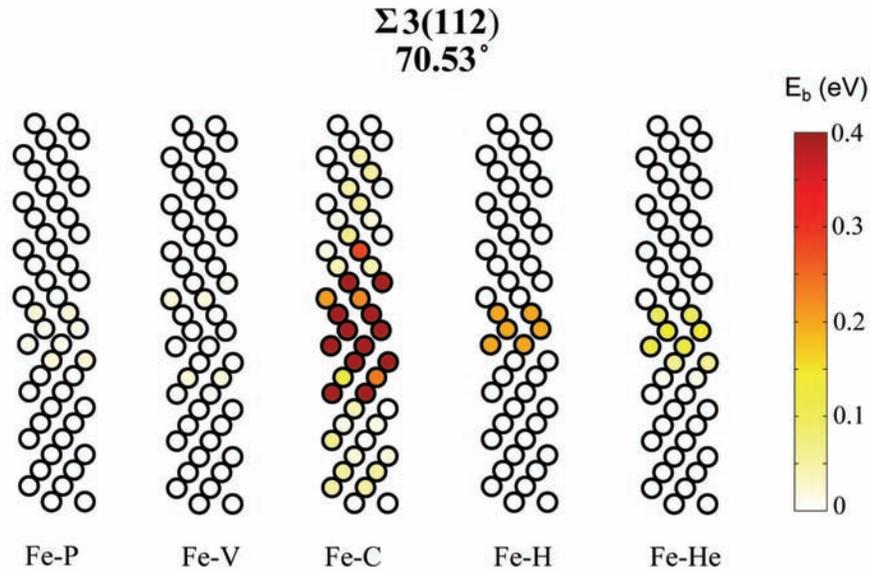

Figure 5. The binding energies of various impurity atoms in various sites for the Σ3(112)θ=70.53° boundary.

Binding energies corresponding to another high angle <110> boundary, the Σ11(332)θ=129.52° GB, is depicted in Figure 6. A few observations arise: (1) Fe-V shows very low binding energies similar to the Figure 5, (2) Fe-H exhibits a slightly higher binding energy as compared to Fe-P for the Σ11(332) GB, followed by Fe-P and Fe-He, and (3) Fe-C exhibits the highest binding indicating that C is highly favorable to segregate to the Σ11(332) GB. These results indicate that the propensity of an impurity atom to bind to a GB depends not only on the interactions between the local GB atoms but also the specific binding site. Probing the binding behavior for a variety of boundaries can provide a more fundamental understanding of the relationship between the binding energetics and atomic structure, which is critical for design applications.

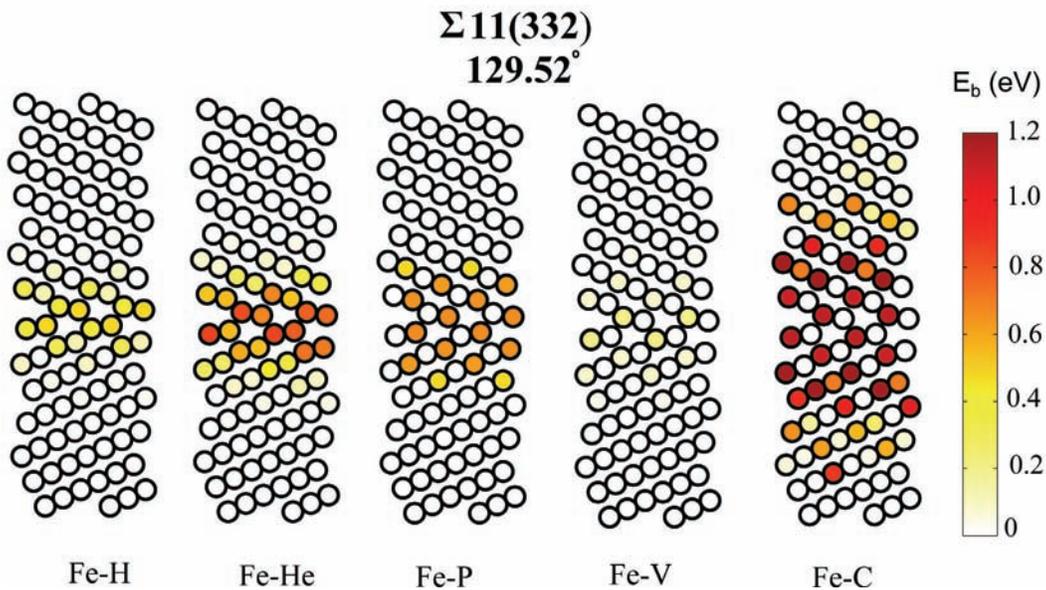

Figure 6. The binding energies of various impurity atoms in various sites for the Σ11(332)θ=129.52° grain boundary.

Next, we investigated the variation of binding characteristics of a high angle <100> symmetric tilt Fe GB as a function of the distance from the center of the GB plane. This global mapping can identify the predominant length scale corresponding to the GB-affected region and also the trend with respect to the binding energies for impurities at GBs. Here, we probed the variation with respect to one substitutional atom (P) and one interstitial atom (H) over the entire length scale. These elements are highly embrittling in nature [10,69] and this mapping also provides information about the symmetry and distribution of binding energies for the system. Several sites within 15 Å of the Σ13(320)θ=67.38° GB are selected for calculating these binding energies The respective plots are shown in Figure 7. It is evident that (1) the variation of energies is highly symmetrical about the GB center (0 Å), (2) the length scale associated with binding energies approaching that of the bulk is about 10 Å for Fe-P and 8 Å for Fe-H system, (3) the maximum binding energy corresponds to approximately 0.55 eV for Fe-H and 0.85 eV for Fe-P, and (4) the majority of sites are energetically favorable for segregation, except for a few sites that behave like the bulk lattice. Therefore, the binding ability of impurity atoms to a GB can be evaluated by assessing the spatial (site to site) and global (length scale) variations in binding energies.

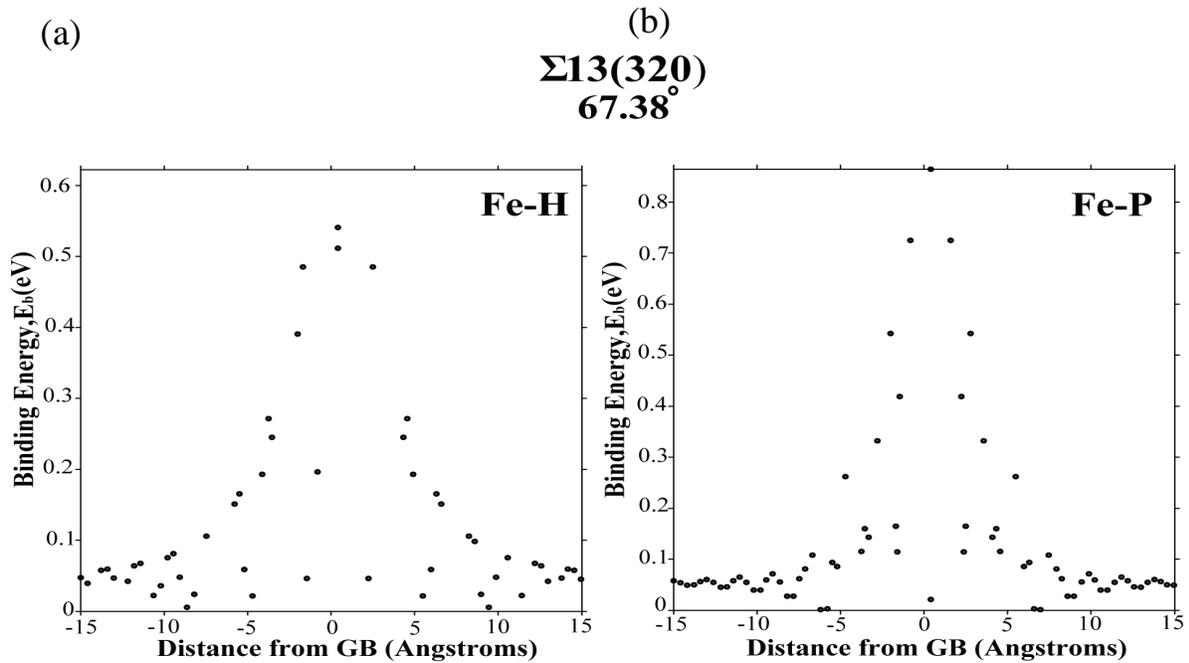

Figure 7. The variation of binding energy as a function of GB distance for the Σ13(320)θ=67.38° GB with (a)H at the interstitial site; and (b) P at the substitutional site. The energies for both the cases converge to the bulk binding energy at distances away from the GB center with H and P exhibiting a maximum binding energy of 0.55 eV and 0.85 eV respectively.

In summary, this methodology provides a representation of point defects and impurity atoms segregation/binding while taking into account the different GB structures. The segregation of solute atoms can alter the cohesive property of boundaries by reducing GB and free surface energies.

3.3 Cohesive energy

The embrittlement strength associated with a boundary and an impurity atom is dictated by the GB cohesive strength. In other words, the cohesion of impurity elements to the GB is influenced by the degree of segregation at the GB and free surface (FS) region. As previously discussed, the preference to segregate to a particular site is determined by the segregation (or binding) energy. Here, the $\Sigma 5(210)\theta=53.13°$ symmetric tilt boundary was chosen to study the impact of segregation energy on the change in cohesive energy. Multiple calculations are performed to investigate the segregation behavior of the two substitutional elements (V and P) at the GB and FS. For all GB systems, as the site where the element is placed moves from GB region to bulk, the segregation energies converge to 0 eV for both the P and V cases (refer to the binding energy plots, Figures 3-7). The segregation energies at the GB region and the FS region are selected based on the position of the site and represented in Figure 8. This plot indicates that (1) there exist sites in the GB where it is energetically favorable for P and V to segregate and (2) it is also energetically favorable to segregate to the free surfaces. The determination of the cohesive strength can also lead to identifying elements that are beneficial for the boundary and those that induce embrittlement. When the GB segregation energy is higher than that of the FS, then the impurity atoms are termed as a cohesion reducer as it reduces the GB cohesive energy, which ultimately leads to the embrittlement of the GB. For cases where the FS segregation energies are significantly higher than the GB, embrittlement is not energetically favored and the impurity is termed a cohesion enhancer. In the case of Fe-P, the FS segregation energy is slightly lower than that of the GB (~ 5.12%), indicating the GB has higher embrittlement potency due to P segregation. On the other hand, in the case of Fe-V for the same boundary, the GB segregation energy is lower than the FS segregation energy, indicating that V does not contribute towards GB embrittlement.

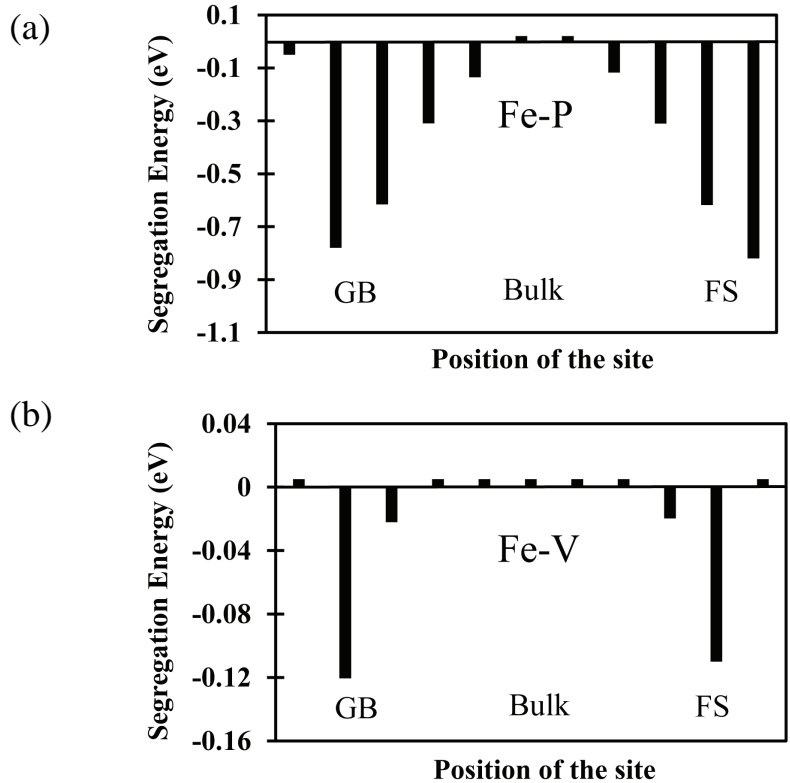

Figure 8. Variation in segregation energies for at the GB and FS for Σ5(210) θ=53.13° GB with (a) P at the substitutional site, and (b) V at the substitutional site. In the case of Fe-P, GB segregation energy is higher than at the FS whereas for the Fe-V GB segregation energy is lower than at the FS indicating that V segregation is beneficial to the boundary.

The susceptibility to embrittlement phenomena was also evaluated by the change in cohesive energy of the GB due to the segregation of impurity atoms. The cohesive energy for the pristine boundary was calculated using Eq. (2) to be 4.94 $J/m^2$. This cohesive energy can also be viewed as the energy per unit area that must be supplied to separate the boundaries. In the case of Fe-P, the cohesive energy of the boundary with one impurity atom is about 4.58 $J/m^2$, i.e., the cohesive energy associated with the segregation of one P atom leads to a decrease of about 0.34 $J/m^2$ as compared to the pure boundary. In the case of Fe-V, the cohesive energy of the boundary with one impurity increases to 5.17 $J/m^2$ which translates into an increase in the cohesive energy of the boundary of approximately 0.59 $J/m^2$. This behavior shows that V has a significant beneficial effect on Fe grain boundary cohesion, while P has a detrimental effect on grain boundary cohesion. Thus, this atomistic assessment of local and global characteristics of impurity atoms at the GB provides a framework that can be extended to a variety of boundaries to determine the binding characteristics and cohesion capability of elements that are either intentionally (alloying) or unintentionally (impurity) added to the matrix material. This energetic analysis can be applied to determine the favorable alloying elements and GB structures to develop materials safe from embrittlement attacks in the presence of stress.

## Conclusions

In this study, molecular statics simulations were employed to analyze the energetics associated with various elements (helium, hydrogen, carbon, phosphorous, and vanadium) at either substitutional or interstitial positions to the four <100> (Σ5 and Σ13 GBs) and six <110> (Σ3, Σ9, Σ11 GBs) symmetric tilt grain boundaries in alpha-Fe. The segregation and binding characteristics were probed site by site as well as a function of distance from the GB. The results show that the binding characteristics depend on the local arrangement of the atoms at the GB as well as the impurity-host atom interaction which also determines the degree of binding. There are significant sites within the GB region where binding energies are positive and all the energies converge to the bulk value as the site moves from GB to bulk. It is also found that the Σ3(112) coherent twin GB exhibits lower binding characteristics as compared to the other boundaries examined herein. The GB cohesive strength was investigated for the Σ5(210)θ=53.13° GB with two elements: P and V. The energy calculations indicate that V has a beneficial effect on the GB cohesion whereas P has a detrimental effect on the GB cohesion. Thus, assessing the local and global binding behavior of an impurity atom to the GBs is useful for designing materials which are suitable for extreme environment applications.

## Acknowledgement

The authors would like to recognize Dr. W. Mullins and Dr. A.K. Vasudevan from the Office of Naval Research for providing their insights and valuable suggestions. This material is based upon work supported by the Office of Naval Research under contract No. N000141110793.